%% file: paper_metallicity.tex
\documentclass[usenatbib]{mn2e}
\usepackage{appendix}
\usepackage{graphicx}
\usepackage{float}
\usepackage{cite}
\usepackage{amssymb,amsmath}
\usepackage[mathscr]{eucal}
\usepackage[usenames]{color}

\input{macro}

\begin{document}

\title[Gas and Metal Contents of Galaxies] 
{Gas and Metal Contents of Galaxies and Gaseous Halos: Preventive versus Ejective Feedback}
\author[] {
Yu Lu$^{1}$\thanks{E-mail: luyu@carnegiescience.edu}, 
H.J. Mo$^{2}$,
Zhankui Lu$^{2}$
  \\
  $^1$ Observatories, Carnegie Institution for Science, 813 Santa Barbara Street, Pasadena, CA 91101, USA
  \\
  $^2$ Department of Astronomy, University of Massachusetts, Amherst
  MA 01003-9305, USA
  }
 

\date{}

\maketitle

\begin{abstract}
Using a semi-analytical approach we investigate the characteristics of predictions 
for the masses and metallicities of the baryonic matter in and around galaxies
made by three galaxy formation models. These models represent three 
different feedback scenarios: one model with purely ejective feedback,  one model with 
ejective feedback with reincorporation of ejected gas, and one preventative model. 
We find that, when the model parameters are adjusted to predict the correct stellar masses 
for a range of halo masses  between $10^{10}$ to $10^{12}\msun$, these
three scenarios have very different predictions for the masses and metallicities 
of the interstellar and circum-galactic media. Compared with current observational 
data, the model implementing preventative feedback has a large freedom to 
match a broad range of observational data, while the ejective models 
have difficulties to match a number of observational constraints simultaneously, 
independent of how the ejection and reincorporation are implemented.  
Our results suggest that the feedback process which regulates 
the amounts of stars and cold gas in low-mass galaxies is preventative in nature.   
\end{abstract}


\section{Introduction}\label{sec:introduction}

One of the most fundamental questions in galaxy formation is how baryons 
are accreted into and expelled from dark matter halos as galaxies form. 
Using statistical approaches, a number of studies have 
revealed a picture in which the baryonic mass locked into stars and cold 
gas is a strong function of galaxy mass \citep{Yang2003a, Behroozi2010a, Moster2013a, Papastergis2012a, Lu2014c}. 
The Milky Way sized galaxies at the present day 
have the largest baryon mass fraction in stars, and this fraction decreases 
rapidly as the galaxy mass increases and decreases.  
How such a relation is established is one of the 
most important questions in galaxy formation. A widely adopted 
galaxy formation scenario 
attributes the low baryon mass fraction in low-mass halos to strong galactic 
outflows generated by the feedback of star formation 
 \citep[e.g.][]{Dekel1986a, Bower2006a, Croton2006a}. 
On the other hand, a number of authors have realized that preventing 
baryons from ever collapsing into dark matter halos may also be 
a viable solution \citep{Mo2002a,Benson2003d, Mo2004b,
van-den-Bosch2003a, Oh2003a, Mo2005a, Lu2007a, Lu2015a}.
Clearly, how feedback works in reality becomes a central question in 
galaxy formation theory. In this paper, we present a comparative 
study of three representative feedback models to understand the 
impact of ejective versus preventive feedback scenarios on their predictions for 
the masses and metallicities in different components of disk galaxies. 

The details of feedback processes 
are still poorly understood. In the commonly adopted ejective scenario, 
feedback of star formation is assumed to produce strong outflows which 
can expel a large amount of gas mass from the host halo 
\citep[e.g.][]{Benson2003a, Somerville2008a, Guo2011c}. However, 
the properties of the outflows, such as their masses and velocities, 
and how they are coupled to the halo gas are still uncertain.  
More recently, a more sophisticated scenario, in which ejected gas is allowed
to reincorporate back to the host halo after a certain time scale is 
proposed \citep[e.g.][]{Henriques2013a}. In addition to 
the same uncertainties that the simple ejective model has, the ejective 
model with reincorporation also has uncertainties in the 
time scale for the gas reincorporation. The physics of these assumed processes 
is still poorly understood. In the preventive scenario, 
some early feedback is assumed to change the thermal state of the 
intergalactic medium around dark matter halos so that a fraction of baryons 
is prevented from collapsing into dark matter halos in the first place. 
For example, as shown in \citet{Lu2015a}, when the circum-halo 
medium is preheated by early feedback to a certain level of entropy, the 
thermal pressure of the preheated gas can support the gas against 
gravitational pull, resulting in a low baryon mass fraction for low-mass halos.
As demonstrated in \citet{Lu2015a}, such a model seems to be able to reproduce 
the observed stellar and cold gas 
mass fractions, sizes of disk galaxies, and star formation histories 
of galaxies with mass comparable to or lower than the Milky Way, 
without invoking any strong outflow as adopted in the ejective scenario.    
An important question is then whether these different feedback scenarios
can be distinguished with current observational data. 

Metallicities in different phases of a galaxy is expected to put interesting constraints 
on feedback models. Metals are produced by star formation and their distribution
is affected by feedback. If, for example, the feedback drives weak outflows, 
metals would be found only in regions where star formation has
taken place. On the other hand, if strong outflows are driven to 
large distances, metals produced by star formation can be 
redistributed not only in ISM (cold gas) but also
in the hot halo gas and even in the intergalactic medium. 
In this paper, we adopt a semi-analytic galaxy formation model introduced 
in \citet{Lu2015a} to better understand how the masses and metallicities of 
baryonic matter in different phases are affected by different implementations 
of feedback. The goal is to examine how current and future observations
may  be used to distinguish between different feedback scenarios. 

The paper is organized as follows. 
In \S\,\ref{sec:model}, we describe the models adopted in this paper. 
We present our model predictions and comparisons with observations 
for the evolution of the baryonic masses in different components 
in \S\,\ref{sec:mass}, and for the metallicities in different phases 
in \S\,\ref{sec:metal}. Our conclusions and the discussion of the 
implications of the results are presented  in \S\,\ref{sec:conclusion}.
Throughout the paper, we use a $\Lambda$CDM cosmology with 
$\Omega_{\rm M,0} = 0.27$, $\Omega_{\Lambda,0} = 0.73$, 
$\Omega_{\rm B,0} = 0.044$, $h = 0.70$, $n = 0.95$, and $\sigma_8 = 0.82$. 

\section{Models}\label{sec:model}

We adopt the semi-analytical model (SAM) introduced in \citet{Lu2015a} and focus 
on different implementations of star formation feedback. To do this, we fix every other 
parts of the model, but vary the recipe for star formation feedback to study 
the impact of feedback on the observational consequences of the masses and metallicities 
in stars, ISM and circum-galactic medium (CGM). We concentrate on galaxies 
with mass comparable to or smaller than that of the Milky Way.
We consider three representative models, which differ from each other by 
their assumptions on how feedback governs gas inflow and outflow. 

The first model is a typical ejective feedback model presented in \citet{Lu2015a}.
This model,  referred to as Model-EJ in the following, captures the essence of 
many ejective feedback models in the literature in terms of how outflows driven by star 
formation feedback affect the baryon contents of low-mass galaxies 
\citep[e.g.][]{Bower2006a, Kang2006a, Somerville2008a}. 
Halos are assumed to accrete baryons at a rate equal to the halo mass 
accretion rate multiplied by the universal baryon fraction $f_{\rm b}=\Omega_{\rm B,0}/\Omega_{\rm M,0}$, 
and a certain fraction of cold gas is assumed to be ejected out of the host 
halo as stars form. The ejected mass is assumed to be proportional to the 
star formation rate with a mass-loading factor that varies with 
the halo virial velocity $\vvir$. Based on the energy conservation argument, 
the mass-loading factor is assumed to be inversely proportional to $\vvir^2$ 
and the mass ejection rate is written as 
\begin{equation}
\dot{M}_{\rm ej}=\psi \,\alpha_{\rm LD} \left({200 {\rm km/s}\over \vvir}\right)^2,
\end{equation}
where $\psi$ is the star formation rate (SFR), and $\alpha_{\rm LD}$ is the 
normalization of the mass-loading factor at $\vvir=200\,{\rm km/s}$.
The ejected mass is assumed to leave the halo and is added into an 
`ejected mass' component. For the model considered here,  we assume that 
the ejected mass never comes back to the halo to explore the maximum 
effect of the ejective feedback scenario. Based on these assumptions, the mass of the 
hot halo gas is given by  
\begin{equation}
M_{\rm hot} = f_{\rm b} \mvir - M_* -M_{\rm cold} - M_{\rm ej},
\end{equation}
where $\mvir$ is the virial mass of the halo, $M_*$ is the stellar mass 
and $M_{\rm cold}$ is the cold gas mass of the galaxy. 
The normalization factor, $\alpha_{\rm LD}$, is tuned so that the model 
prediction matches the stellar mass of Milky Way sized halos as
derived from recent empirical models that match galaxy 
stellar mass functions at multiple redshifts \citep{Behroozi2013a, Lu2014e}. 
We find that $\alpha_{\rm LD}=1$ provides a reasonable fit to both the 
stellar mass and cold gas mass in Milky Way sized halos. 
Since the mass loading factor scales as $\vvir^{-2}$, 
lower-mass halos have larger mass-loading factors. 
Moreover, in order to explain the metal mass in the circum-galactic 
medium (CGM), we deposit a fraction, $\alpha_{\rm Z}$, of the metal yield from star formation 
directly into the hot halo gas, and the rest is mixed uniformly 
into the ISM before being ejected out of the halo. We explore the 
effect of changing the fraction of this direct metal injection into the 
halo gas and found that $\alpha_{\rm Z} =0.1$ produces a good match to the 
OVI gas mass observed in COS-Halos, assuming the OVI ionization fraction $f_{\rm OIV}=0.2$ \citep{Tumlinson2011a}. 
A lower level of direct metal injection would predict a too low oxygen mass to be consistent with 
the observational constraints. 

The second model considered here is an extended ejective model, in which 
the ejected gas mass is allowed to reincorporate into the halo hot gas after a period of time. 
Specifically, we implement the model proposed by \citet{Henriques2013a}, 
where the gas reincorporation rate is written as 
\begin{equation}
\dot{M}_{\rm ej} = - {M_{\rm ej} \over t_{\rm reinc}},
\end{equation}
where $t_{\rm reinc}$ is the time scale for the ejected gas to fall back 
into the halo. \citet{Henriques2013a} utilized a MCMC method to optimize a flexible 
version of the L-Galaxies SAM to match the evolution of galaxy luminosity 
function, and found the following model for  $t_{\rm reinc}$:
\begin{equation}
t_{\rm reinc}=\gamma {10^{10}\msun \over \mvir},
\end{equation}
where $\gamma$ is a free parameter tuned to be 
$\gamma=1.8\times10^{10}\,{\rm yr}$. To be consistent with the 
Henriques model,  the prescription and parameter values of the outflow model
we adopt follow \citet{Henriques2013a} exactly. 
Specifically, a fraction of cold gas is assumed to be 
heated and deposited into the hot halo gas reservoir and 
another fraction is assumed to be ejected and leave the halo. 
The ejected gas is then subject to reincorporation. The details of the 
model parameterization can be found in \citet{Guo2011c},  and the 
detailed modeling  of the reheating, ejection and reincorporation 
can be found in \citet{Henriques2013a}. In what follows, this model 
is referred to as Model-RI. Because this model already has 
reheated ISM to be mixed with the halo gas, we do not assume 
any additional channel of metal injection into the hot halo.   

The third model is the preventative feedback model introduced in \citet{Lu2015a}. 
In this model, the effect of feedback is not to eject gas out of halos 
but to prevent a fraction of the baryonic matter from collapsing into halos.
This is achieved by assuming that the intergalactic medium (IGM) is 
pre-heated to a certain entropy level so that a fraction of baryonic mass will 
not be able to collapse into the shallow gravitational potential wells of low-mass halos 
owing to the thermal pressure of the preheated gas. 
As demonstrated with hydrodynamical simulations in \citet{Lu2007a}, a uniform 
entropy distribution of the IGM results in the fraction of baryon 
mass that collapses into a halo to be proportional to the halo mass squared, 
$M_{\rm b}\propto M_{\rm vir}^2$, which is needed to match the shallow 
low-mass end slopes of local galaxy stellar mass and HI-mass functions 
\citep{Mo2005a}. In addition, the halo gas in this model is assumed to establish 
a hot corona with an extended central core in its density profile. 
\citet{Lu2015a} showed that this preventative scenario can reproduce the 
observed cold gas mass, stellar mass,  star formation history, and size 
for galaxies with masses comparable to or smaller than that of the Milky Way 
without invoking any outflow after preheating.  On the other hand, however, such a 
universal prevention cannot suppress star formation sufficiently 
for more massive galaxies at late time. Thus, a ``late quenching" process 
is needed for these galaxies. Here we consider a mechanism in which the stars 
and/or the central black hole in the central galaxy produce certain amounts 
of heating to offset the cooling of the halo gas. Motivated by the results of 
\citet{Lu2014f} that the classical bulge mass and central 
black-hole mass in a galaxy with total stellar mass 
$10^{10}{\rm M}_\odot <M_*< 10^{11}{\rm M}_\odot$ both scale
roughly as $M_*^2$,  we assume that the heating rate depends 
on $M_*$ as 
\begin{equation}
\dot{H} = \epsilon \left({M_*\over 10^{10}\msun}\right)^2 \msun {\rm yr}^{-1}\, ,
\end{equation}
where the heating efficiency is set to be $\epsilon=5$. 
In every time step, we calculate the halo cooling rate, $\dot{C}$, as 
described in \citet{Lu2015a}, and the heating rate.  The net cooling rate is then 
\begin{equation}
\dot{M}_{\rm cool} = {\rm max}[\dot{C} - \dot{H}, 0]\,.
\end{equation}
Moreover, we also allow a moderate level of outflow, assuming a constant mass-loading 
factor of unity for all galaxies. This assumption is consistent with directly observations of outflows
\citep[e.g.][]{Bouche2012a, Kacprzak2014a},  and is also allowed by statistical inference 
from an empirical model of galaxy evolution constrained by stellar mass functions 
at different redshifts \citep{Lu2014d}.  This model is referred to as Model-PR in the 
following.  

For all the three models, we follow not only the mass of baryonic matter in each 
phase as in \citet{Lu2015a}, but also metallicities in the different phases. 
For all the phases considered here, namely the halo gas, cold gas, ejected 
gas, and stars, we only track the average metallicity for each phase, 
but do not follow the details of the metallicity distribution. 
Throughout the paper, we adopt the solar metallicity as 
$[12 + {\rm log(O/H)}]_{\odot} = 8.69$ \citep{Allende-Prieto2001a, Asplund2009a} 
and $Z_{\odot} = 0.0134$ \citep{Asplund2009a}, and the total metal yield 
is taken to be $p=0.03$, which is defined as the ratio between the mass of newly produced metals that are ejected into 
the ISM and the gas mass turned into stars. With the assumption that all galaxies have the 
same  abundance ratio as the Sun,  the oxygen yield is $p_{\rm oxy}=0.44 p$.

\section{Baryonic masses in different components}
\label{sec:mass}

\begin{figure*}
\centering
\includegraphics[width=0.9\textwidth]{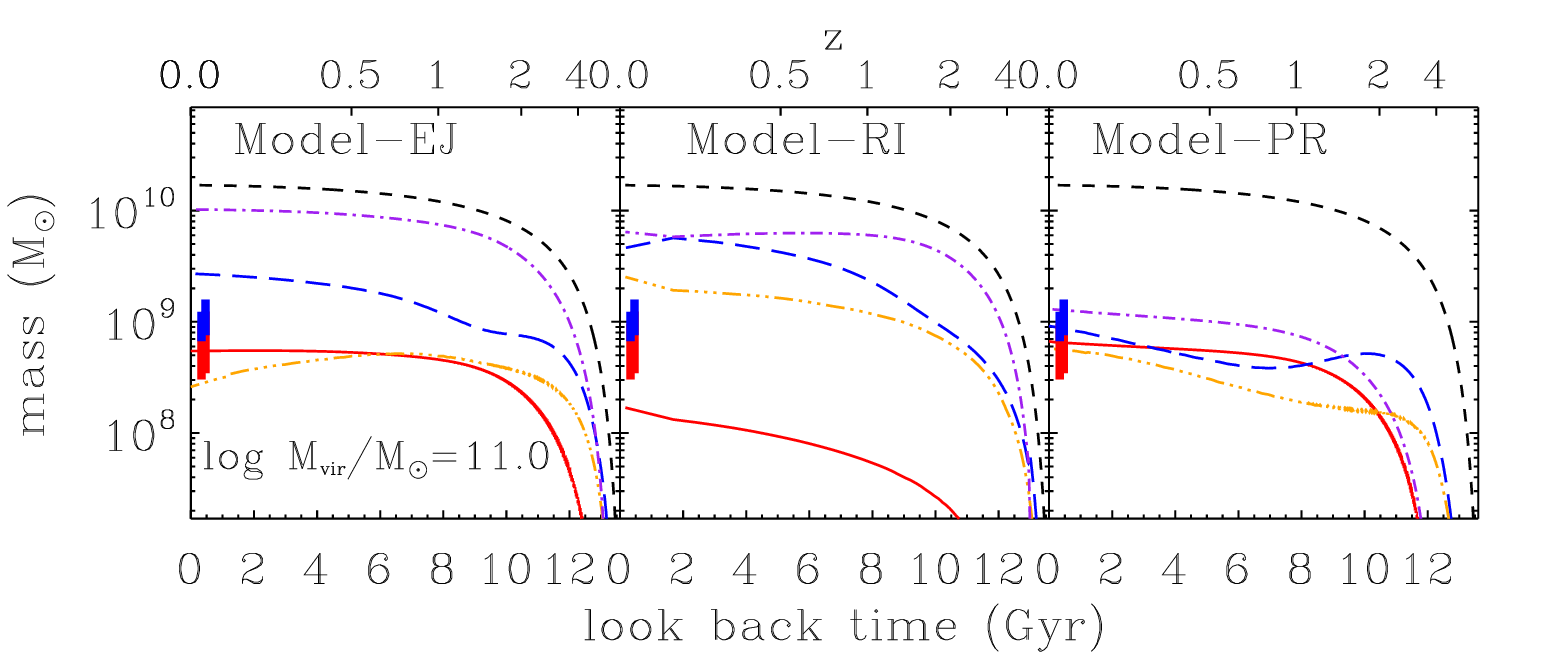}
\includegraphics[width=0.9\textwidth]{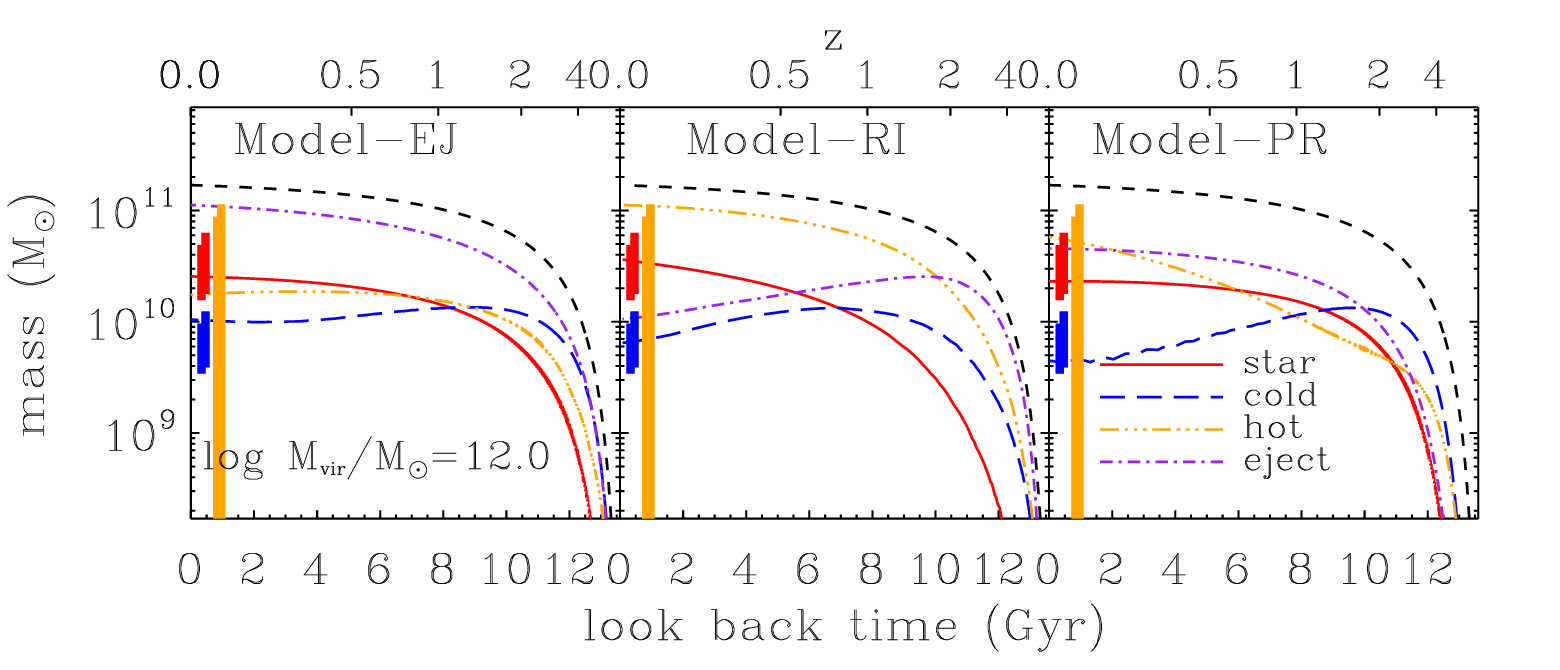}
\caption{
The mass of different baryonic component as a function of time (redshift) 
for halos with a final mass of $10^{12}\msun$ and $10^{11}\msun$ at the present day.
The predictions of different models are labeled by their names in different panels. 
The red and blue lines show the stellar mass and the cold gas mass in model galaxies, 
respectively. The orange line is the gas mass in the hot halo, and the purple line 
is the mass ejected out of the halo. The black dashed line plots the halo mass 
multiplied by the cosmic baryon mass fraction $f_{\rm b}$. The vertical bars mark the 
ranges of observational results.  For stellar mass, the observational data constraint 
is taken from \citet{Behroozi2013a} and \citet{Lu2014e}.
For the cold gas, the observational constraint is from \citet{Papastergis2012a}. 
The constraint for the hot halo gas is adopted by bracketing data constraints 
obtained from \citet{Fang2006a} and \citet{Bregman2007a}.
}
\label{fig:mass_hist}
\end{figure*}

\begin{figure*}
\centering
\includegraphics[width=0.9\textwidth]{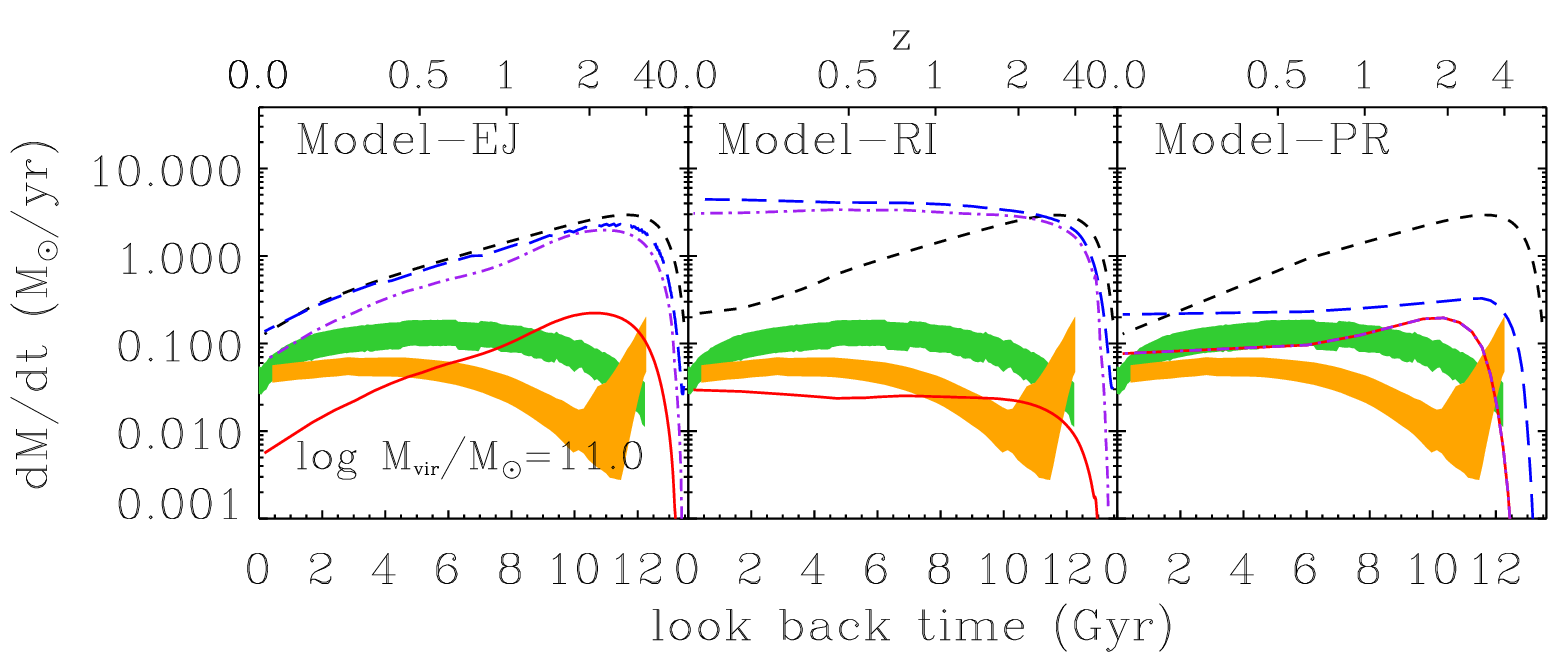}
\includegraphics[width=0.9\textwidth]{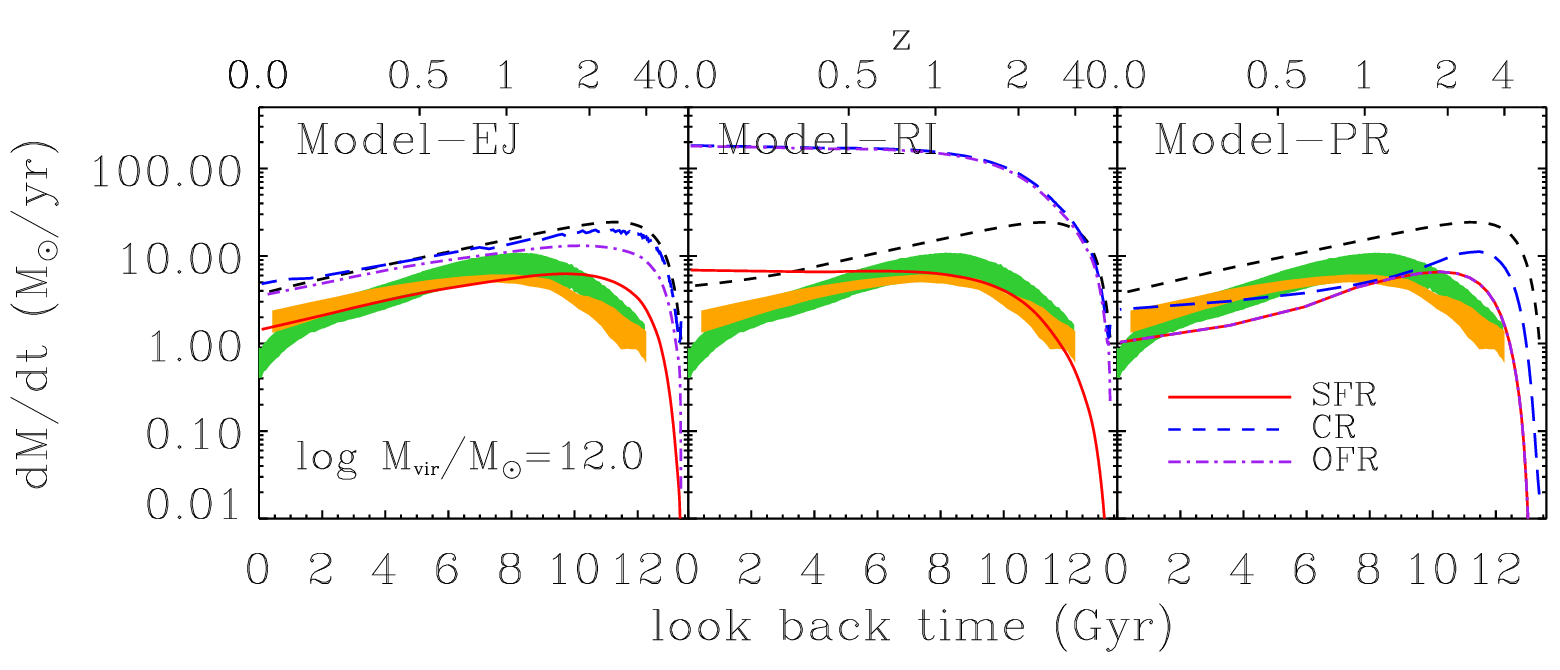}
\caption{The rates of baryonic mass conversion among different phases as functions 
of time (redshift) for halos with a final mass of $10^{12}\msun$ and $10^{11}\msun$ at the present day.
The predictions of different models are labeled by their names in different panels. 
The red line is the star formation rate; the blue line is 
the radiative cooling rate of the hot halo; the purple line is the outflow rate. 
The black dashed line denotes the halo mass accretion rate multiplied by the 
universal baryon fraction, $f_{b,0}=0.17$.  
For comparison, the star formation rates inferred by \citet{Behroozi2013a} 
and \citet{Lu2014e} using empirical models are over-plotted
as the green and orange bands, respectively.  
}
\label{fig:rate_hist}
\end{figure*}

To understand how the baryonic masses in stars, cold gas, hot halo gas and ejected 
components evolve with time in different models, we apply these models to typical halo 
mass accretion histories (MAHs) of two final halo masses, $10^{11}\msun$ 
and $10^{12}\msun$. For simplicity, we adopt the fitting model of \citet{McBride2009a}, 
\begin{equation}\label{equ:mah}
M_{\rm halo}(z) = M_{\rm halo,0} \left( 1+z\right)^{\alpha} \exp(-\eta z),
\end{equation}
where the normalization $M_{\rm halo, 0}$ is the halo mass at $z=0$ and 
the parameters $\alpha$ and $\eta$ determine the shape of the MAH. 
This functional form can be considered as a generalization of two other widely 
adopted models proposed earlier by \citet{Wechsler2002a}
and \citet{van-den-Bosch2002a}, and has been shown to accurately describe
the assembly histories over a wide range of halo masses. It is also similar to other 
models proposed in the literature \citep{Zhao2003a, Zhao2003b, Zhao2009a, Dekel2009a}.  
\citet{McBride2009a} showed that the parameters $\alpha$ and $\eta$ have 
broad distributions. For our model, we choose the typical values of these two 
parameters: $\alpha=0.6$ and $\eta=0.9$ for $M_{\rm halo, 0}=10^{12}\msun$ 
halos, and $\alpha=0.8$ and $\eta=0.9$ for $M_{\rm halo,0}=10^{11}\msun$. 
This model matches well the simulated MAH, as shown in  \citet[][]{Lu2015a}. 

Following the spirit of SAM, we compute, at each time step, the halo gas accretion, 
hot gas distribution, radiative cooling, star formation and feedback, and keep 
track of the evolutions of the masses in stars, cold gas in the disk, hot gas in the halo, 
and the gas ejected out of the halo. The model predictions are shown in 
Figure\,\ref{fig:mass_hist},  with the upper row for 
$M_{\rm halo,0}= 10^{11}\msun$ halos, and 
the lower row for $M_{\rm halo,0}=10^{12}\msun$. 
While the three models predict similar albeit 
distinguishable histories for the stellar mass and cold gas mass, 
they predict very different histories for the halo gas and ejected gas masses. 
Model-EJ predicts the ejection of a large amount of gas and the ejected 
mass is kept out of the halo. Consequently the majority (more 
than half) of the baryonic mass originally associated with the halo is 
predicted to be ejected for both the halo masses.  
For $M_{\rm halo,0}=10^{11}\msun$, the cold gas is the second dominating 
component, but for $M_{\rm halo,0}=10^{12}\msun$, hot halo gas mass is
comparable to the mass of the stellar component while the cold gas mass 
is the smallest of all. The ejected mass never decreases with time
because of the absence of re-incorporation.  
The situation is very different in Model-RI. Although this 
model also implements a strong outflow, the ejected mass is re-incorporated 
back into the host halo at late times. Owing to the increasing reincorporation 
timescale with decreasing halo mass, the ejected gas outside the halo is always 
the dominating fraction for low-mass halos, but it is a significantly lower fraction for 
$M_{\rm halo,0}=10^{12}\msun$. Furthermore, for the 
$10^{12}\msun$ halos in this model, the ejected mass dominates 
the baryon mass budget only at very early time ($z>2$), 
and this fraction decreases rapidly at low redshifts as the halo mass increases 
to the regime where re-incorporation becomes important.
Finally, for Model-PR, the total baryon mass budget is governed 
by the preheating entropy, which causes low-mass halos to be relatively poor 
in baryons.  Since a low mass-loading factor of outflow is assumed 
in this model, the ejected mass simply follows the stellar mass history with a delay 
caused stellar evolution \citep[see][for details]{Lu2015a}. 
For $M_{\rm halo,0}=10^{11}\msun$, the cold gas mass is about $10^9\msun$ 
at the present day, and the stellar mass is about half of the cold gas mass. 
For $M_{\rm halo,0}=10^{12}\msun$,  on the other hand, the stellar mass 
reaches about $3\times10^{10}\msun$ while the cold gas mass is about one 
order of magnitude lower. For both $M_{\rm halo,0}=10^{11}\msun$
and $10^{12}\msun$, the hot halo gas mass is comparable to the stellar 
mass at the present time. 

We compare the predicted stellar and cold gas masses at the present day with  
the results of \citet{Behroozi2013a}, \citet{Lu2014e} and 
\citet{Papastergis2012a}, shown as the vertical red and blue bars. 
The observational results for the hot halo gas 
are from \citet{Fang2006a}, \citet{Bregman2007a}, \citet{Yao2008a}, 
\citet{Tumlinson2011a} and \citet{Gupta2012a}, and shown as the vertical 
yellow bars. Because all the models are tuned to match the present-day 
stellar mass and cold gas mass of Milky-Way sized halos, it is not 
surprising that all models reproduce well the observed 
stellar and cold gas masses for $10^{12}\msun$ halos at 
$z=0$. The measurements of hot halo gas mass have large uncertainties.
Both Model-EJ and Model-PR can accommodate the observational results 
comfortably,  while the prediction of Model-RI seems to be too high.

For $M_{\rm halo,0}=10^{11}\msun$, both Model-EJ and Model-PR can achieve 
an excellent agreement with data for the final stellar mass, 
but Model-RI under-predicts the final stellar mass by a factor of 2 to 3. 
Even assuming strong outflows, both Model-EJ and Model-RI over-predict 
the cold gas mass by a factor of few. This over-prediction stems from  
the fact that a large amount of cold gas is settled in 
an extended gaseous disk where star formation is low 
due to the low surface density and feedback is not efficient to 
get rid of the cold gas. If we allow the gas disks to be smaller so that 
star formation can occur in a larger fraction of the disk, the predicted gas 
mass can be reduced. However, as shown in \citet{Lu2015a},
to match the cold gas mass requires disk sizes to be much smaller than 
real disks.  
We also note that the overproduction of cold gas mass 
causes an underprediction of gas-phase metallicity (see next section for details). 
When the gas-phase metallicity is sufficiently low, the star formation rate surface density 
strongly depends on the gas surface density. This nonlinear dependence causes an upturn 
in stellar mass in the final 2 Gyrs for $10^{11}\msun$ halo.  
In contrast to the ejective models, Model-PR can reproduce the stellar mass and 
cold gas mass for both $10^{12}\msun$ and $10^{11}\msun$ halos. 
In this model,  the amount of cold gas settled in the outer parts of disks is 
smaller, as a large fraction of the gas is not able to make to the disk because 
of preheating.   

\begin{figure*}
\includegraphics[width=0.9\textwidth]{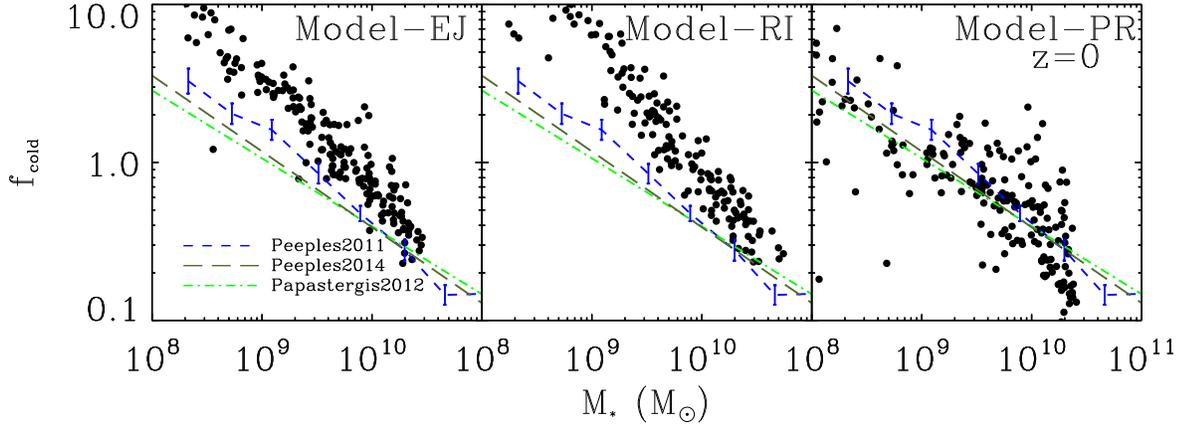}
\caption{Cold gas mass - stellar mass ratio as a function of galaxy stellar mass at $z=0$ predicted by 
different models, as labeled by their names in different panels. The mean relations derived 
from observational data by \citet{Peeples2011a}, \citet{Peeples2014a}, and \citet{Papastergis2012a} 
are shown by different lines as noted in the figure. 
}
\label{fig:fcold}
\end{figure*}

Figure \ref{fig:rate_hist} shows the rates at which the baryonic masses in different components 
transfer between each other. We focus on the cooling rate of halo gas, star formation rate of the cold gas, 
and the outflow rate as functions of time.  For the outflow rate, we plot the rate at which the gas is leaving the galaxy. 
For Model-EJ, the outflow gas goes out of the halo, but for Model-RI a fraction of the outflow is 
retained in the halo.  For simplicity, we do not distinguish between outflows
going into the IGM and those going into the CGM. Since mass ejection is powered by star 
formation feedback in all the three models, the outflow rate is proportional to the SFR.
The difference between models is in the assumed mass-loading factor. 
In Model-EJ, the mass-loading factor is proportional to $\vvir^{-2}$, and so
low-mass halos have much higher outflow rate per star formation rate. 
Model-RI adopts a very large mass-loading factor to balance the gas re-incorporated into the 
gaseous halo, and so both the cooling rate and ejection rate are at least one order 
of magnitude higher than the SFR.  In Model-PR, the cooling rate deviates significantly from the halo 
accretion rate because of preheating, and the star formation is largely controlled 
by the reduced rate of gas accretion. For comparison, we include the SFR as a function of 
time for both $M_{\rm halo,0}=10^{11}\msun$ and $10^{12}\msun$, as inferred   
by \citet{Behroozi2013a} and \citet{Lu2014e} from observational data.  All the three models 
reproduce the average star formation history of $10^{12}\msun$ halos quite well.
For $M_{\rm halo,0}=10^{11}\msun$, Model-PR matches the empirical results well; 
Model-RI moderately under-predicts the SFR at low redshift; Model-EJ fails completely:
it predicts too high a SFR at high $z$ and too low a SFR at low $z$.     

Figure \ref{fig:fcold} shows the present-day cold gas to stellar mass 
ratio $f_{\rm cold}$ as a function of galaxy stellar mass predicted by the three models. 
The predictions are made by applying the models to a random set of halo mass 
accretion histories selected from the Bolshoi cosmological $N$-body simulation \citep{Klypin2011a}
for the mass range between $10^{10}$ and $10^{12}\msun$. 
As one can see, the gas fractions predicted by all the three models for high 
mass halos are consistent with observational results. This again is 
because the model parameters are all tuned to match the properties 
of Milky Way sized galaxies. Both of the ejective models (Model-EJ and Model-RI) 
predict a steep decline of the gas fraction with stellar mass, so that 
both models significantly over-predict the gas fractions in low-mass galaxies.
Varying the normalization of the mass-loading factor in the ejective
models only shifts the amplitude of the predicted relation but has
negligible effect on the slope of the relation, and so cannot fix the 
mismatch. In contrast, Model-PR reproduces the observed $f_{\rm cold}$--$M_*$ 
relation quite well. This model predicts a large scatter in the cold gas fraction 
for high-mass galaxies. This is largely caused by the `late quenching' we 
introduced, in which the suppression of gas cooling depends strongly on stellar 
mass. Also, star formation in halos less massive than $10^{11}\msun$ 
depends sensitively on halo mass accretion history in this model.
Depending on whether a halo has acquired most of its mass 
before or after the onset of preheating, the total amount of stars that can 
form can be very different. This has the effect of increasing the scatter 
in the predicted $f_{\rm cold}$--$M_*$ relation.

\section{Metal Masses and Metallicities in Different Gas Components}\label{sec:metal}

\begin{figure*}
\centering
\includegraphics[width=0.9\textwidth]{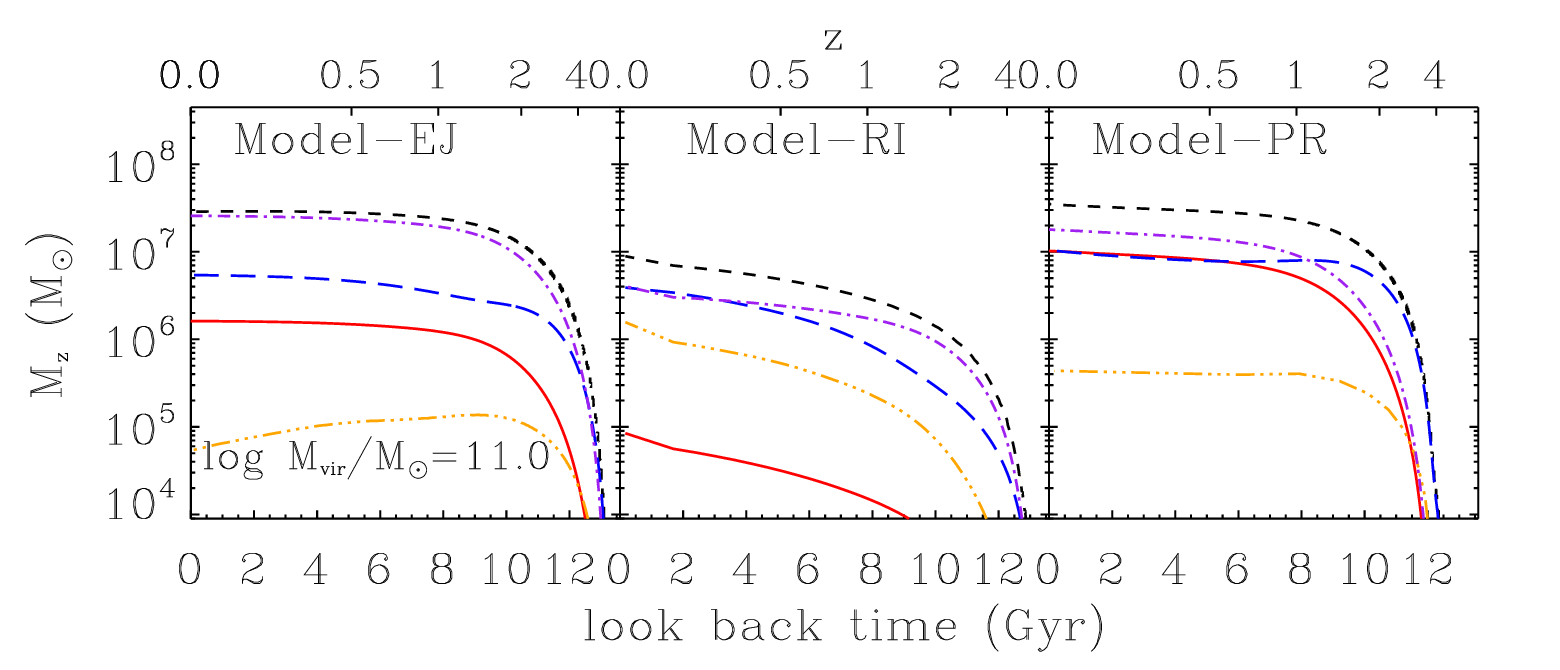}
\includegraphics[width=0.9\textwidth]{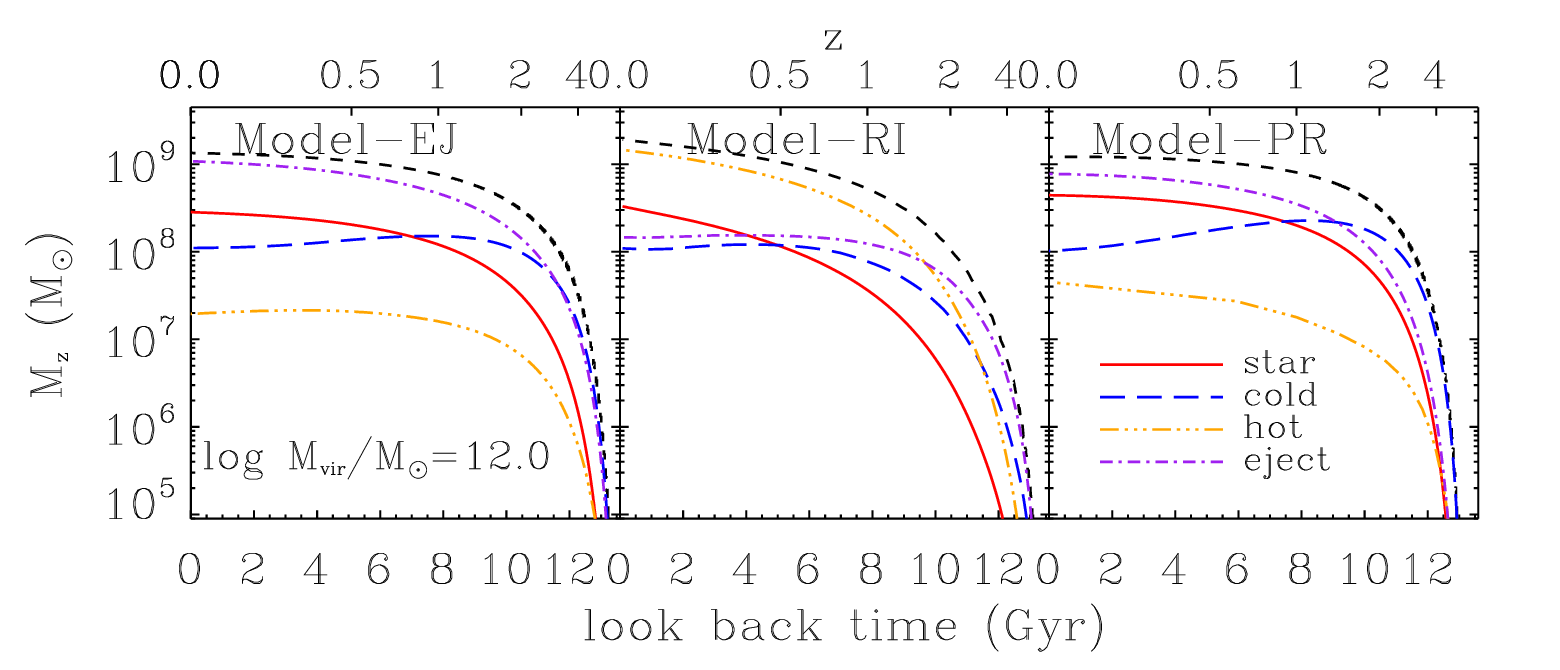}
\caption{The metal masses in different phases of baryonic matter as 
predicted by the three models (EJ, RI and PR, as indicated in the panels)
for halos with final masses of $10^{12}\msun$ (lower panels) and  $10^{11}\msun$ 
(upper panels) at the present day. The red lines are for the metal masses locked in stars, 
the blue lines for metals in cold gas, the orange lines for metals in gaseous halo, 
and the purple lines for metals that are ejected out of halo. The black dashed lines
denote the metal mass available for the galaxy, which is approximated by $p{M_* \over {1-R}}$.
}
\label{fig:z_hist}
\end{figure*}

We also follow the histories of the metal mass in stars, cold gas, hot halo gas, and ejected gas 
of model galaxies hosted by $10^{11}\msun$ and $10^{12}\msun$ halos. 
we adopt the usual instantaneous recycling approximation for metals produced by start formation 
to be returned to ISM and CGM.  
The time delays between the nuclear synthesis of different heavy elements are ignored for simplicity. 
This approximation is supported by the results of \citet{Peeples2014a}, who showed that 
the Oxygen mass, which is mainly produced by Type II SNe, is roughly a constant fraction 
of the total metal mass over the stellar mass range of our study. 
Figure \ref{fig:z_hist} shows the histories of metal mass in each component as a
function of time (redshift) predicted by the three models. 
In each panel, we also show the total available metal mass, $pM_*/(1-R)$
(with $p$ being the metal yield and $R$ the return fraction), as star formation 
proceeds. We adopt $R=0.43$, which is proper for a Chabreir IMF.
The ejective feedback in Model-EJ expels the enriched gas from the galaxy,
and so most of the metal mass ends up in the ejected gas. 
In Model-RI, the ejected mass returns back to the halo after a short time delay 
for the Milky Way sized halo, but can stay in the halo for a longer 
period of time. Consequently,  most of the metal mass in such a halo
is in the hot halo gas. For $10^{11}\msun$ halos,  on the other hand, 
the ejected component contains a larger fraction of metals than the halo 
gas, because gas re-incorporation is slow for low-mass halos
and much of the ejected gas stays outside of the halo.  
Furthermore, since low-mass halos are predicted to have a high cold gas fraction, 
the cold gas in such a halo is predicted to contain a large fraction of the total 
metal mass in Model-RI.  In contrast, Model-PR predicts that the metal 
masses locked in stars and cold gas are always a substantial fraction
of the total, while the metal mass in the halo gas is lower by a factor of 
a few.  

\begin{figure*}
\centering
\includegraphics[width=0.9\textwidth]{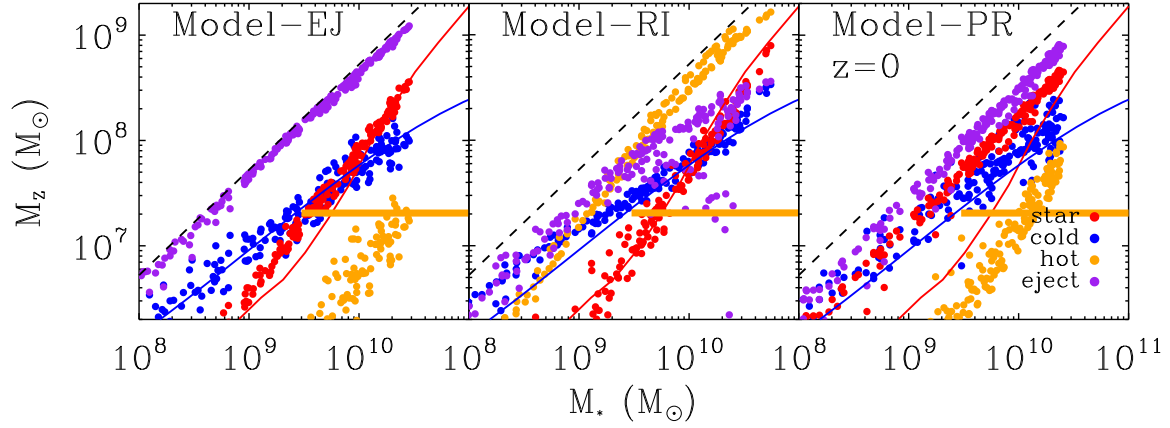}
\caption{The relation between metal mass in different baryonic mass components and 
the stellar mass for model galaxies predicted by different models. The model name is noted in 
each panel. The red dots denote metal mass locked in stars, the blue dots denote metal mass 
in cold gas in galaxies, and orange dots denote metal mass in hot halo, and the purple dots 
denote metal mass in gas that is ejected out of the halos. The horizontal orange bar marks the 
observational estimate of the metal mass in halo gas around galaxies with masses over the range 
of stellar masses covered by the bar as measured in COS-Halos through OVI line \citep{Tumlinson2011a}. 
Solar oxygen abundance pattern is assumed. 
}
\label{fig:mmetal}
\end{figure*}

In Figure \ref{fig:mmetal} we show the metal masses in different phases
versus  stellar mass for galaxies at $z=0$ predicted by the three 
models, and compare the model predictions with the compilation of 
observational data by \citet{Peeples2014a}. In addition, we also include 
a horizontal bar to represent the observational estimate of metal mass 
in hot halos derived from OVI absorptions of COS-Halos samples, assuming $p_{oxy}=0.44p$ 
and the same ionization fraction $f_{\rm OVI}=0.2$ as adopted in \citet{Tumlinson2011a}. 
All the models predict that the metal mass increases with stellar mass, except 
for the ejected mass predicted by Model-RI. The general increasing trend is set by 
the total metal mass available, which is plotted as the black dashed line in the figure.  
Galaxies with higher stellar masses can produce more metals, but where the metals 
are located depends strongly on the feedback model.  

In Model-EJ, a substantial fraction of the ISM is blown out of galaxies, and because 
all the ejected mass is assumed to leave the halo forever, most of the metal mass is
predicted in the ejected component for galaxies of different masses. 
The model also has a small fraction of metals in the hot gas, which stems 
from our assumption that a fraction of metal yield is deposited in the halo. 
For the particular parameter ($\alpha_{\rm Z}=0.1$) assumed here, the predicted 
metal mass in the hot halo is about $2\times10^7\msun$ for galaxies with final 
stellar masses $\sim 10^{10}\msun$, which is consistent with the 
observational results \citep{Tumlinson2011a}. The predicted metal mass in the cold gas component 
also matches well the observational data presented in \citet{Peeples2014a}. 
We note that increasing $\alpha_{\rm Z}$ can increase the metal mass in the hot gas to 
make the model prediction agree with the data better, but it will
lower the metal mass in the cold gas component so that the agreement 
with the observational results does not hold anymore.  Overall, this model nicely 
reproduces the metal masses as functions of galaxy stellar mass in all
components,  except the ejected component, for which data constraint
is not available. 

Model-RI predicts a trend of the metal mass in stars 
and cold gas that is similar to that in Model-EJ 
due to the common assumption of ejective feedback. However, the two models
predict very different trends for the metal mass in the hot gas and the ejected gas.
Since the reincorporation is fast for high-mass galaxies \citep{Henriques2013a}, 
a large fraction of metals can return back to the hot halo. Consequently,  
the predicted metal mass in the ejected component first increases with 
increasing stellar mass, and then decreases, for galaxies with stellar 
masses $>10^{10}\msun$. The metal mass in the halo gas predicted by this model 
increases rapidly with increasing  galaxy mass owing to both the increase of total metal 
production and the halo mass-dependent reincorporation. 
Furthermore, since a large fraction of metals stays in the halo, the model 
predicts high metal mass in the cold phase,  especially for low-mass halos
where cooling from the halo gas is efficient. 

Model-PR adopts a moderate ejection with a mass-loading factor equal to unity, 
and Fig.\,\ref{fig:mmetal} shows that even with such a small loading factor 
the ejected mass can still contain a large fraction of all the produced metals. 
Most of the metals that are not ejected are predicted to be locked in stars and 
cold disk gas. As in Model-RI, metals in the hot halo gas are due to the assumed 
direct deposition in the halo. Again, we see that for the particular choice of 
$\alpha_{\rm Z}=0.1$, the model predicts about $10^7\msun$ of metals in 
the halo of a galaxy with $M_*=10^{10}\msun$, which is consistent with the 
observational results. The predicted metal mass  in cold gas matches 
the observation result, but the model seems to over-predict the metal mass 
in stars. As we will discuss in the following section, this discrepancy is the only 
main problem for this model, and suggests that the assumption of uniform 
mixing of metals in the ISM may be invalid.

\begin{figure*}
\includegraphics[width=0.9\textwidth]{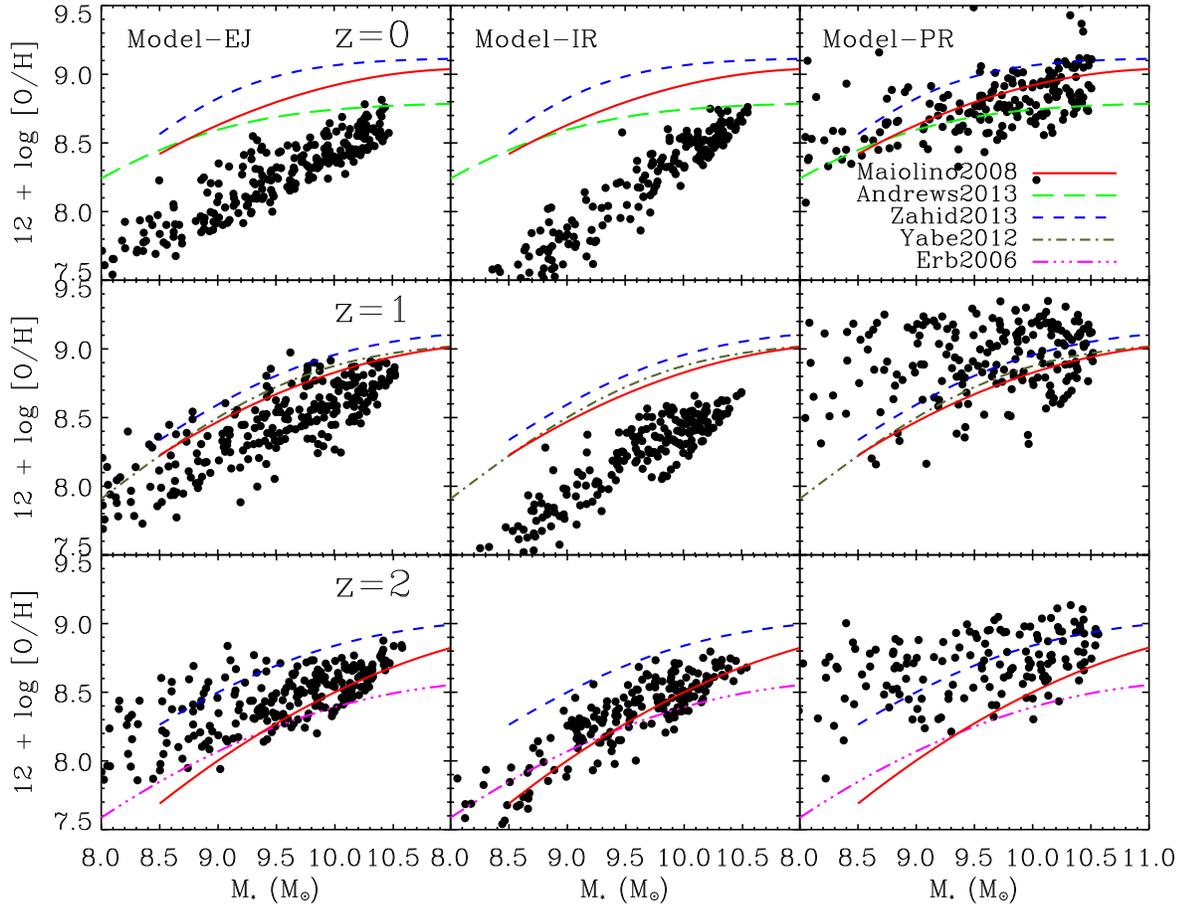}
\caption{The metallicity of the cold gas (ISM) as a function of stellar mass at 
$z =$ 0, 0.5, 1, and 2 predicted by Model-EJ, Model-IR, and Model-PR.
The fitting functions of the mean relations derived from the observational data of 
\citet{Andrews2013a, Maiolino2008a}, \citet{Zahid2013a}, \citet{Yabe2012a} and \citet{Erb2006a}
are shown in each corresponding panel for comparison.  
}
\label{fig:cgmetal}
\end{figure*}
  
The relation between the gas-phase metallicity and galaxy stellar mass 
has been studied observationally over the redshift range $0<z<3$ by many authors 
\citep[e.g.]{Tremonti2004a, Erb2006a, Maiolino2008a}. However, hydrodynamic 
simulations and semi-analytic models of galaxy formation have difficulty in 
reproducing the relation over the redshift range probed by the observation
\citep{De-Lucia2004a, de-Rossi2007a, Mouhcine2008a}. 
In Figure \ref{fig:cgmetal} we show the predictions of our three models 
for the oxygen metallicity in cold disk gas at three different redshifts, 
$z = 0$, 1, and 2. For comparison, we plot three sets of observational results
obtained by  \citet{Andrews2013a}, \citet{Maiolino2008a}, \citet{Zahid2013a}, 
\citet{Yabe2012a} and \citet{Erb2006a}. 
The ejective feedback model (Model-EJ)
predicts only a weak evolution in the oxygen metallicity - stellar 
mass relation from $z=2$ to the present day, which is at odd with 
the increase of the metallicity with time for a fixed stellar mass 
shown in the observational data 
\citep{Mannucci2009a, Maiolino2008a, Mannucci2010a, Zahid2013a}. 
The predicted metallicities at low redshifts are too low  to match 
the observation. The reason for this is twofold. First, the model over-predicts 
the cold gas fraction for a given stellar mass, which leads to a reduced metallicity for the gas. 
Second, metals are also ejected out of galaxies with outflow, further reducing 
the metallicity. The under-prediction and the lack of evolution also appear 
in Model-IR.  \citet{Henriques2013a} found that their reincorporation model 
can match  the {\it stellar} metallicity - stellar mass relation for local 
galaxies, assuming a metal yield $p=0.047$ that is 60\% higher than 
what is assumed here. However, our results show that the reincorporation model
is not able to reproduce the evolution in the gas phase metallicity unless
$p$ is time-dependent.  The preheating model Model-PR predicts 
a moderate level of evolution and a large scatter in the gas phase metallicity for given stellar mass, 
and the model predictions are  consistent with the observational data assuming $p=0.03$. 

\begin{figure*}
\includegraphics[width=0.9\textwidth]{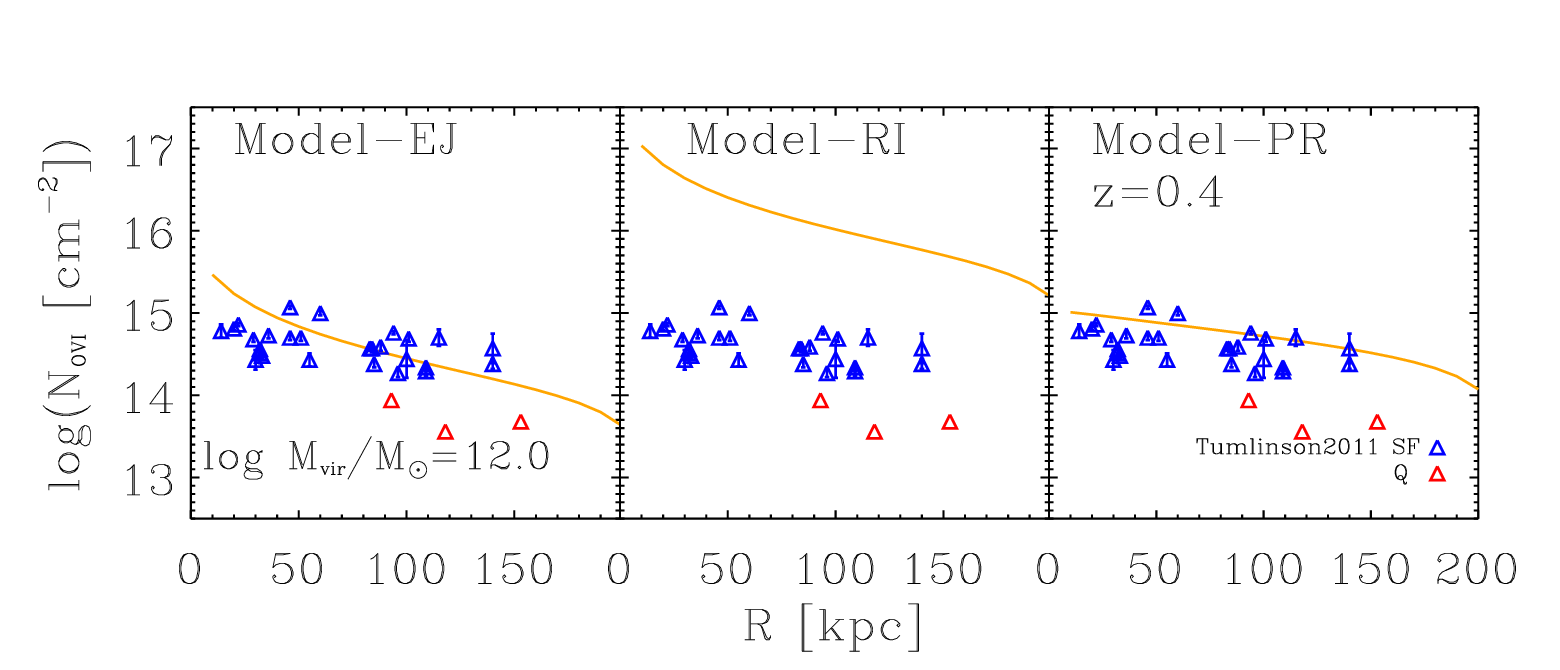}
\caption{The column density of OVI as a function of impact parameter (projected radius from galaxy center) for 
a $10^{12}\msun$ halo predicted by different models. The blue data points are observational data 
for star forming galaxies and the red data points are for quiescent galaxies obtained in \citet{Tumlinson2011a}. 
}
\label{fig:Novi_profile}
\end{figure*}

Since we assume that metals in the halo gas is perfectly mixed in a 
density profile, we can make predictions for the column density of 
any chemical species as a function of the impact parameter 
(distance from galaxy center). The COS-Halos survey has measured the 
OVI column density profile for a sample of $L^*$ galaxies around redshift $z=0.4$. 
Here we make the corresponding predictions for a $10^{12}\msun$ halo 
at $z=0.4$ and compare our model predictions with the data.
In the calculation, we follow \citet{Tumlinson2011a} and assume that the ionization 
fraction of OVI,  $f_{\rm OVI}=0.2$.  As shown in Fig.\ref{fig:Novi_profile}, 
all the three models can roughly reproduce the shape of the column density 
profile. The predictions of Model-EJ and Model-PR agree with the data reasonably well, 
but the column density profile predicted by Model-RI is about 2 orders of magnitude 
too high. This over-prediction by Model-RI is again due to the reincorporation of the 
ejected gas, which is enriched in metal.

\begin{figure*}
\includegraphics[width=0.9\textwidth]{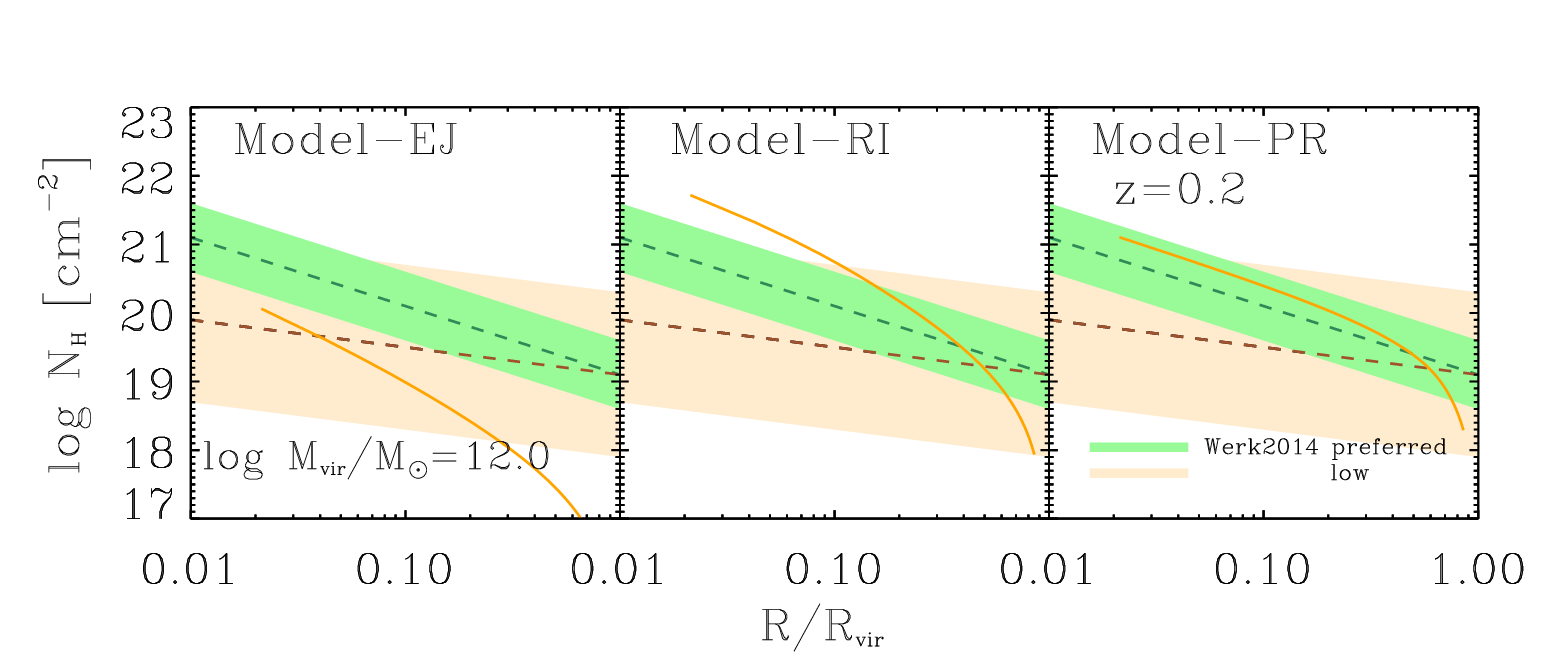}
\caption{The cooling gas column density as a function of impact parameter (projected radius from galaxy center) for 
a $10^{12}\msun$ halo predicted by different models. The green and orange bands represent the 
`preferred' and lower limits of the observational estimates of \citet{Werk2014a}.
}
\label{fig:coolgas_profile}
\end{figure*}
   
\citet{Werk2014a} have estimated the mass and column density profile 
of cool gas in the circum-galactic medium for 44 Milky Way sized galaxies in 
the COS-Halos sample.  They found that the total amount of cool gas 
(with temperature in the range $10^4\,{\rm K}<T<10^5\,{\rm K}$) is larger than 
about $6.5\times10^{10}\msun$, and  that the cool gas column density profile 
can be well described by a shallow power-law with a power index $\sim -1$ or lower.  
Assuming the cooling halo gas in our model to be this cool component in the 
circum-galactic medium, we compare our model predictions for a typical 
$10^{12}\msun$ halo at $z=0.2$ with the observational results.
To do this, we estimate the amount of the halo gas at different radii 
that can cool in a given time interval based on the profile of the hot halo gas. 
The gas cools from a shell at radius $r_i$ covers a $4\pi r_i^2$ area. 
Assuming that the cooled gas falls onto the halo center in one free-fall time from 
the radius with a constant speed, the cool halo gas that drops from radius $r_i$ 
has a radial distribution proportional to ${1 \over r^2}$, where $r<r_i$. 
Thus, the total mass density of the cooled halo gas from all the outer shells at a radius $r$ 
can be written as 
\begin{equation}
\rho_{\rm cool}(r)= {1 \over r^2} \int_r^{\rvir} {\Lambda(r_i) \tau_{\rm ff}(r_i) r_i }  {\rm d}r_i, 
\end{equation}
where $\Lambda(r_i)$ is the cooling rate of the halo gas per unit volume at $r_i$, and $\tau_{\rm ff}$ is the free-fall time of the 
gas element from the radius where it cools. 
The free-fall time is estimated as 
\begin{equation}
\tau_{\rm ff} = 14.4{\rm Gyr} \left({r_i \over 1{\rm Mpc}}\right)^{3 \over 2} \left[{M(<r_i) \over {\rm 10^{12}\msun}}\right]^{-{1 \over 2}}\,,
\end{equation}
where $M(<r_i)$ is the total mass (dark matter and baryon) within radius $r_i$. 
We then calculate the column density from the radial profile and 
the cool gas column density profiles predicted by the three
different models, which are shown in Fig.\,\ref{fig:coolgas_profile}. 
The amplitudes of the column density profiles predicted by both 
Model-EJ and Model-PR agree with the observational results, while 
that predicted by Model-IR seems too high. 
These are consistent with the predicted cooling rates shown in Fig.\,\ref{fig:rate_hist}:
Model-EJ and Model-PR predict a much lower rate than Model-IR. 
More interestingly, both ejective feedback models predict profiles that are 
significantly steeper than the observational estimates.
In contrast, the preheating model predicts an extended cooling gas profile
similar to the observational results. 

\section{Discussion and summary}\label{sec:conclusion}

\setcounter{table}{0}
\begin{table*}
\begin{minipage}{115mm}

\begin{center}
\begin{tabular}{| l | l | c | c | c | c |} \hline

Mass Scale   &  Quantity      &      Model-EJ &  Model-RI  & Model-PR   &   Figure \\ \hline

$\mvir=10^{11}\msun$      &     $M_*(z=0) $      &     OK      & too low   &   OK      &      1 \\
      &      $M_{\rm cold}$(z=0)   &  too high  & too high  &   OK    &        1 \\
      &      SFH   & too low at low-z  &  too low at low-z   &   OK     &       2 \\
      &      $f_{\rm cold} (z=0)$     & too high &  too high  &   OK      &      3 \\
\hline
$\mvir=10^{12}\msun$      &      $M_*(z=0)$     &       OK     &  OK        &   OK     &       1 \\
      &      $M_{\rm cold}$(z=0)  &     OK   &    OK     &      OK    &        1 \\
      &      SFH   &  OK     &  OK     &      OK        &    2 \\
      &      $f_{\rm cold} (z=0)$       & OK    &   OK     &      OK       &     3 \\
\hline

$M_*=10^8-10^{10.5}\msun$             & $Z_{\rm ISM}(z=0)$        & too low  &  too low   &   OK       &     6 \\
             & $Z_{\rm ISM}(z=2)$        &  OK   &     OK    &      OK       &     6 \\
             & $M_{\rm Z,*}(z=0)$       & OK    &    OK      &   too high   &     5 \\
             & $M_{\rm Z,hot}$          &  OK   &    too high  &   OK        &    5,7 \\ \hline

\end{tabular}
\label{tab:compare}
\caption{A summary of the success and failure of the three models based on 
comparisons with different observational data. The first column lists the masses of 
model halos or galaxies. The second column lists the quantities in the comparison. 
The third to fifth columns list the performances of the three models. The last column 
lists the corresponding figures from which the conclusion is drawn. 
}
\end{center}
\end{minipage}
\end{table*}

Using a set of semi-analytic models of galaxy formation, we have demonstrated that the gas and 
metal contents of the galaxy population provide important constraints on the feedback 
processes assumed in galaxy formation models.  In particular, the masses and the 
metallicities of baryonic matter in different phases can be used to distinguish two broad 
classes of feedback scenario, namely ejective and preventative feedback. 
We have found that when the models are tuned to predict a correct stellar mass for 
a given halo mass, different feedback mechanisms work differently in redistributing 
the metal-enriched material, and hence predict very different trends for metallicities.  

We have considered three representative models. Two of them implement ejective 
feedback, in which halos accrete baryonic matter together with dark matter at the 
cosmic baryon fraction. To match the observed stellar mass fraction of galaxies in the
ejective scenario, a significant fraction of baryonic matter that cools 
into the central galaxy has to be ejected out of the galaxy by star formation feedback. 
One of the ejective models implements a simple assumption that the ejected material never 
comes back, while the other one considers a more sophisticated model in which 
the ejected gas can be re-incorporated into the halo. The third model implements a 
preventative feedback model, in which feedback 
works to preheat the IGM, and the thermal pressure prevents a fraction of baryon 
from collapsing into low-mass halos. To explore the characteristics of these 
models, we have contrasted their predictions in the masses and 
metallicities of stars, cold disk gas, hot halo gas, and ejected gas. 
Table 1 summarizes the success and failure of the model predictions 
in comparison with existing observational data. In the following we 
discuss the implications of these results.   

First, ejective and preventative models predict different fractions of baryons that 
reside in a halo and outside the halo viral radius. The pure ejective model predicts 
that most of the baryon mass is ejected out of the halo, and only a small fraction 
is retained in the central galaxy and the halo. The ejective model with gas reincorporation 
has similar predictions for low-mass halos, but for Milky Way sized galaxies 
it predicts that the dominating fraction of baryons is in the halo because the 
ejected gas can return quickly back to the halo. 

Second, we have shown that when the outflow mass-loading parameter  is 
tuned to be large enough in the ejective models so that the predicted total 
stellar mass matches the observed stellar mass - halo mass relation, 
model galaxies tend to have too high a gas mass fraction compared 
with observational data. Meanwhile, the ejective models also predict a too 
low gas-phase metallicity. Although stronger outflow with an even larger 
mass-loading factor could reduce the gas fraction in the ejective models, the 
metal mass will inevitably decrease further, resulting in an even lower gas-phase 
metallicity. Thus, the ejective models appear to be unable to accommodate the 
observed gas/stellar mass fraction and gas-phase metallicity simultaneously.  

Third, regardless of the details of how ejection and reincorporation 
are parameterized, the ejective models always predict a 
gas phase metallicity-stellar mass relation with roughly 
fixed amplitude and slope since $z=2$ without strong
time evolution, which is inconsistent with current observations. 

Finally, we have shown that if feedback is primarily ejective and the ejected mass can 
be reincorporated into massive halos rapidly as suggested in recent models 
\citep[e.g.][]{Henriques2013a}, gaseous halos may contain too much metal mass 
to be consistent with COS-Halos observational results. 

In contrast the preventative model can accommodate almost all observational data, 
including the evolution of masses and metallicities of baryonic matter in different phases. 
The only problem is that the model predicts too high metal masses in 
the stelar component of low-mass galaxies. Given that this model predicts the 
correct metal mass in ISM, and the correct total stellar and cold gas masses, 
the problem cannot be solved by assuming stronger outflows with uniform mixing. 
The over-prediction suggests that a significant fraction of stars in low-mass 
galaxies may form in a medium with a significant lower metallicity than the average ISM 
metallicity. This can happen if a fraction of star formation in low-mass 
galaxies occurs in early starburst before $z=2$, when the ISM metallicity has not
 been substantially enriched \citep{Lu2014c}.  Alternatively, a large fraction 
($>70$\%) of the stellar mass in low-mass galaxies may have formed in regions 
where the metallicity is lower than the average. Indeed, as demonstrated in detail in Lu et al. 
(2015, submitted), both the observed stellar-phase metallicity and gas-phase 
metallicity as functions of galaxy stellar mass can be reproduced,  
if star formation occurs in an ISM with inhomogeneous metallicity, 
where a large fraction of the star forming ISM has metallicities 
below the average ISM metallicity. Since real ISM is inhomogeneous 
in metallicity \citep{Smartt1997a, Ho2015a},  this only discrepancy for the preventative model may 
not be a real problem after all. 

In summary, the results we have obtained indicate that the basic assumption of 
the ejective model that all baryonic matter first collapses into a halo 
and the baryonic matter budget in a galaxy is regulated by outflows is 
problematic. This, together with the success of the preventive model considered 
here and in \citet{Lu2015a}, suggests that feedback is 
likely preventative in nature. In the present paper we have explored 
the different models by manually varying the key model parameters. 
These parameters certainly have degeneracy between them. In addition, the two 
processes, ejection and prevention, are expected to be degenerate to some 
extent for some of the observational constraints. It remains to be revealed how 
these two processes are degenerate and what observational data can effectively 
break the degeneracy. We will investigate these questions using the Bayesian 
model inference approach developed by \citet{Lu2011b} in a future paper. 

\section*{Acknowledgement}
The authors thank Andrew Benson, Guillermo Blanc, Mike Fall, Yicheng Guo, 
Rachel Somerville, Risa Wechsler for useful discussions.
YL received partial support from HST-AR-12838, provided by
NASA through a grant from the Space Telescope Science Institute, which is
operated by the Association of Universities for Research in Astronomy,
Inc., under NASA contract NAS5-26555. 
HJM acknowledges the support from NSF AST-1109354.
This work used the Extreme Science and Engineering Discovery Environment (XSEDE), 
which is supported by National Science Foundation grant number ACI-1053575.
\bibliography{/Users/luyu/Documents/bibtex/metallicity}

\end{document}

%% file: macro.tex
\newcommand{\msun}{\>{\rm M_{\odot}}}

\newcommand{\beq}{\begin{equation}}
\newcommand{\eeq}{\end{equation}}

\newcommand{\mvir}{M_{\rm vir}}
\newcommand{\rvir}{R_{\rm vir}}
\newcommand{\vvir}{V_{\rm vir}}

\newdimen\hssize
\newdimen\hdsize 